\documentclass[epj]{webofc}
\usepackage[utf8]{inputenc}
\usepackage[varg]{txfonts}   
\usepackage{booktabs}
\usepackage{xcolor}

\definecolor{darkred}{rgb}{0.4,0.0,0.0}
\definecolor{darkgreen}{rgb}{0.0,0.4,0.0}
\definecolor{darkblue}{rgb}{0.0,0.0,0.4}
\definecolor{abricot}{RGB}{230, 126, 48}
\definecolor{amazon}{RGB}{59,122,87}
\definecolor{bleuimperialfonce}{RGB}{0,65,106}
\definecolor{cerise}{RGB}{222, 49, 99}
\definecolor{caeruleum}{RGB}{53, 122, 183}
\definecolor{violeteminence}{RGB}{108,48,130}
\definecolor{bleunuit}{RGB}{15,  5,  107}
\definecolor{bleuroi}{RGB}{49, 140, 231}
\definecolor{bleuciel}{RGB}{119, 181, 254}

\usepackage[colorlinks,
linkcolor = darkred,
urlcolor  = darkblue,
citecolor = darkgreen]{hyperref}
\usepackage{pgfplots}
\usepackage{tikz}
\usepackage{subfigure}
\usetikzlibrary{pgfplots.groupplots}
\usepgfplotslibrary{fillbetween}
\usepgfplotslibrary{patchplots}

\wocname{EPJ Web of Conferences}
\woctitle{Lattice2017}
\begin{document}
	%
	\selectlanguage{english}
	\title{%
		Thermal Simulations, Open Boundary Conditions and Switches 
	}
	\author{%
		\firstname{Yannis} \lastname{Burnier}\fnsep \and
		\firstname{Adrien} \lastname{Florio}\inst{1}\thanks{Speaker, \email{adrien.florio@epfl.ch}} \and
		\firstname{Olaf}  \lastname{Kaczmarek}\inst{2}\fnsep \and 
		\firstname{Lukas}  \lastname{Mazur}\inst{2}\fnsep
	}
	\institute{%
		Institute of Physics, LPPC, Ecole Polytechnique F\'ed\'erale de Lausanne (EPFL), Lausanne, Switzerland
		\and
		Fakult\"at f\"ur Physik, Universit\"at Bielefeld, D-33615, Bielefeld, Germany
	}
	\abstract{%
		$SU(N)$ gauge theories on compact spaces have a non-trivial vacuum structure characterized by a countable set of topological sectors and their topological charge. In lattice simulations, every topological sector needs to be explored a number of times which reflects its weight in the path integral. Current lattice simulations are impeded by the so-called freezing of the topological charge problem. As the continuum is approached, energy barriers between topological sectors become well defined and the simulations get trapped in a given sector. A possible way out was introduced by L\"uscher and Schaefer using open boundary condition in the time extent. However, this solution cannot be used for thermal simulations, where the time direction is required to be periodic. In this proceedings, we present results obtained using open boundary conditions in space, at non-zero temperature. With these conditions, the topological charge is not quantized and the topological barriers are lifted. A downside of this method are the strong finite-size effects introduced by the boundary conditions. We also present some exploratory results which show how these conditions could be used on an algorithmic level to ”reshuffle” the system and generate periodic configurations with non-zero topological charge.
	}
	
	\maketitle
	\section{Introduction}\label{intro}
	
$SU(N)$ gauge theories on compact and orientable manifolds admit a set of distinct vacua, labeled by an integer topological charge $Q$ \cite{DeWitt1979}. When computing the path integral, all topological sectors need to be included \cite[chap. 23.6]{9780521670548}. Conventional lattice QCD simulations  use periodic boundary conditions; they take place on the torus $\mathbf{T}^4$. Hence, different lattice configurations are characterized by different $Q$'s (for a review of the topology of $SU(N)$ theories on the torus, see \cite{Gonzalez-Arroyo1997}).

As the continuum limit is approached, the topological sectors become well defined \cite{Luscher1982c}, they decouple and small deformations of the configurations do not allow to travel from one sector to the other. The simulations  get stuck in a given sector; this problem is often referred to as "topological freezing" and has been investigated in the literature, see for example \cite{Schaefer2011,Fritzsch2014}.

Different approaches to solve this issue have been or are studied \cite{Luscher2011b,Mages2015a,Czaban2014,Bietenholz2016,thisContriHasen}. In particular, the use of open boundary conditions (OBC) in the time extent has been proposed in \cite{Luscher2011b}. In this context, the base manifold of the simulations is not compact anymore, $Q$ is not quantized  and the topological freezing disappears. An important drawback is the appearance of strong finite-size effects \cite{Mages2015a,thiscontribSommer}.

This contribution is constructed as follow. In section \ref{sec:freezing}, we discuss the topological freezing at non-zero temperature and present quenched results obtained with OBC in the spatial extents.  This is relevant for the computation of the topological susceptibility at non-zero temperatures, which is of utter interest for cosmology and axion physics \cite{thiscontriMoore}. It is also needed for a precise determination  of transport coefficients, whose correlators are known to be sensitive to the topological sampling \cite{Fukaya2014b}. We show that, as expected, the topological charge is not quantized and the topological freezing disappears. However, atop of the topological charge not being anymore a topological invariant, OBC configurations are plagued by strong finite-size effects. 

Thus, in section \ref{sec:switch}, we try to circumvent these problems and present quenched results where OBC are used purely as an algorithmic tool\footnote{A very related idea was presented by M. Hasenbusch \cite{thisContriHasen} at this conference.}. The idea is to generate a certain number of configurations with PBC, then switch to OBC in order to "reshuffle" the system, before going back to PBC. It manages to produce higher topological charge configurations but oversamples the topology. We conclude by discussing outlooks on how this method could be improved.

	\section{Topological Freezing and Spatial OBC}\label{sec:freezing}
	
	To overcome the problem of topological freezing, the idea of using OBC was introduced in \cite{Luscher2011b}. By making the manifold non-compact, one loses the quantization of the topological charge; there are no more distinct topological sectors \cite{DeWitt1979,Luscher2011b}. However, to be able to simulate thermal configurations, one needs periodicity in time \cite{Gross1981}. This enforces the use of OBC in spatial directions. To preserve cubic symmetry, we impose them in all three directions, thus setting\footnote{As explained in the original paper \cite{Luscher2011b}, one also needs to introduce some weights for the boundary plaquettes in the action.}
	
	\begin{equation}
	F_{i\mu}(x)|_{x_i=-a}=F_{i\mu}(x)|_{x_i=N_i a}=0, \ \ \  i=1,\dots,3 \label{eq:OBC}
	\end{equation}
	for a lattice with lattice spacing $a$ and sizes $N_t\times N_x \times N_y \times N_z$.

\begin{figure}
\begin{center}
	\includegraphics{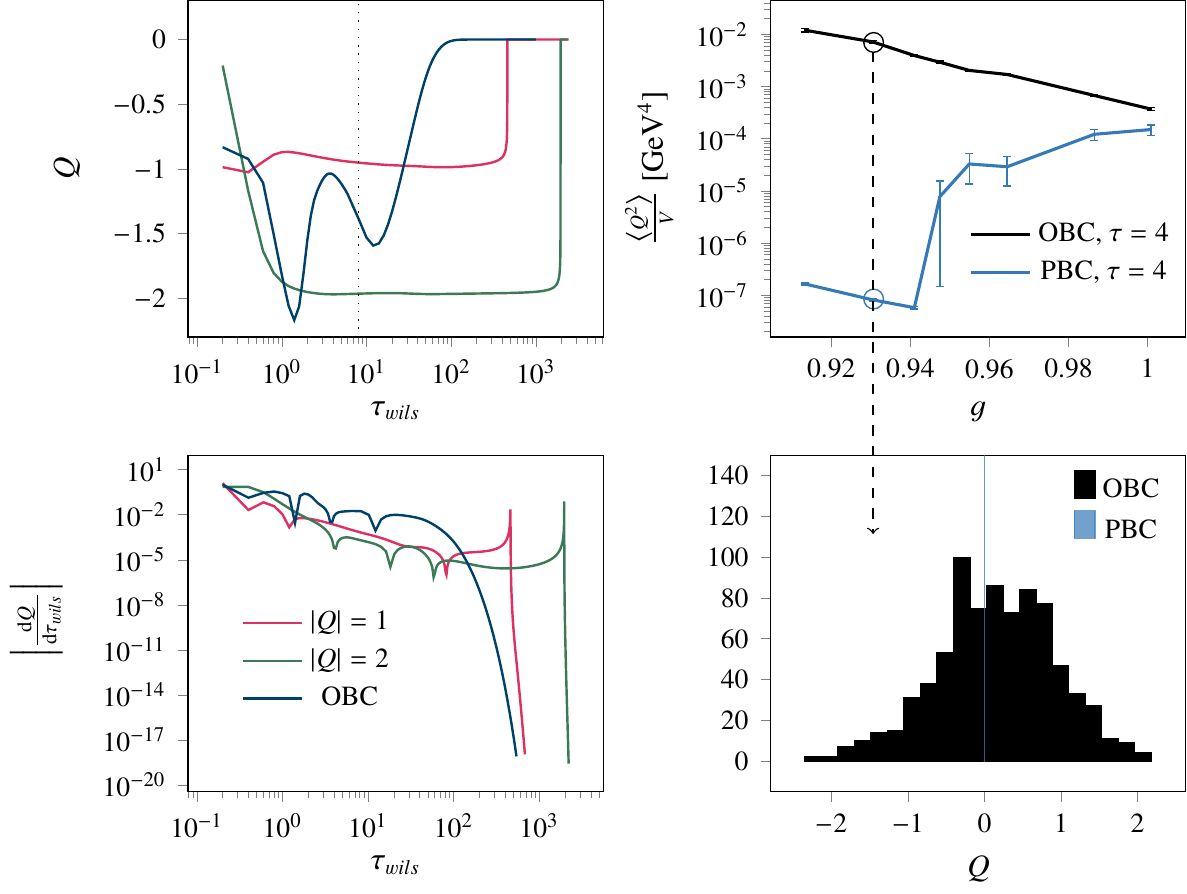}
	\caption[Freezing and Wilson's Flow]{\textbf{Left:} Evolution of the topological charge with the dimensionless flow time, $T=0$. For PBC, it gives rise to a well-defined integer, robust to Wilson's flow evolution. In the case of OBC, there is no clear mean of singling out a flow time at which the topological charge should be defined. The lower plot shows the derivatives of the curve and displays a clear difference between the two types of conditions.
		\textbf{Upper right:} Topological charge variance for different couplings, $T=1.28T_c$. In the case of PBC, we see that the topology freezes. This does not happen in the OBC case. \textbf{Lower right:} Topological charge distributions for OBC (black) and PBC (blue). This clearly illustrates the fact that OBC configurations do not possess an integer topological charge.}
	\label{fig:freezing_wilson}
\end{center}
	\end{figure}

	We start by doing some measurements on $T=0$ configurations, $20^4$ lattices at $\beta=6.2$, where the topological transitions are not damped by the quark-gluon plasma. To compute the topological charge, we use Wilson's flow to evolve the configurations towards the continuum together with the clover discretization of the field-strength tensor. When the continuum theory has distinct topological sectors, the flow dynamically interpolates every configuration to a given sector \cite{Luescher2010}. This  is illustrated by the curves in the left hand-side part of figure~\ref{fig:freezing_wilson}. The upper plot is a semi-log plot of the topological charge's evolution as a function of the dimensionless flow time for given configurations. The topological charge is not quantized at $\tau=0$. First, in the case of PBC (\textcolor{amazon}{green} and \textcolor{cerise}{red} curves), when $\tau$ is increased, the discrete configurations are flown towards continuous bundles. They are representative of different topologies; they carry different integer topological charges. We see that this quantity is unambiguously determined and very robust under flow evolution. The less stable $Q=-1$ charge decays at about $\tau=600$, which corresponds to a smoothing range of roughly $\rho\approx70$ lattice sites ($\rho\approx\sqrt{8\tau}$, see \cite{Luscher2011b}), more than three times the actual time extent of the lattice. The observed fact that the $|Q|=2$ charge is more stable than the $|Q|=1$ is a general feature, which had already been observed on the lattice in \cite{BilsonThompson:2003zi}, using some smearing methods. On the torus, there exists no self-dual configuration for $|Q|<2$, see \cite{Braam1989} and references therein.  The  $|Q|=1$ sectors only have infima of their action but no minima; they are less stable.
	
	On the other hand, when dealing with OBC (\textcolor{bleuimperialfonce}{blue} curves), no plateau is reached; the charge continuously flows towards zero. It reflects the non-existence of distinct topological sectors; this is the feature which allows for a good sampling of the topological fluctuations.
	
	To pursue this idea, we simulated some toy lattices (small aspect ratios and volumes) at $T=1.28T_c$ for different lattice spacings. In the right hand-side part of figure~\ref{fig:freezing_wilson}, we show the ratio of the topological charge variance by the lattice volumes, both for PBC and OBC configurations. For PBC, as the coupling is reduced, the topological susceptibility drops; different sectors are not sampled anymore; the simulation freezes. For OBC, as expected, it does not occur. No topological barriers impede the Monte-Carlo (MC) evolution; the topological fluctuations are well sampled.

		\begin{figure}
			\begin{center}
				\includegraphics{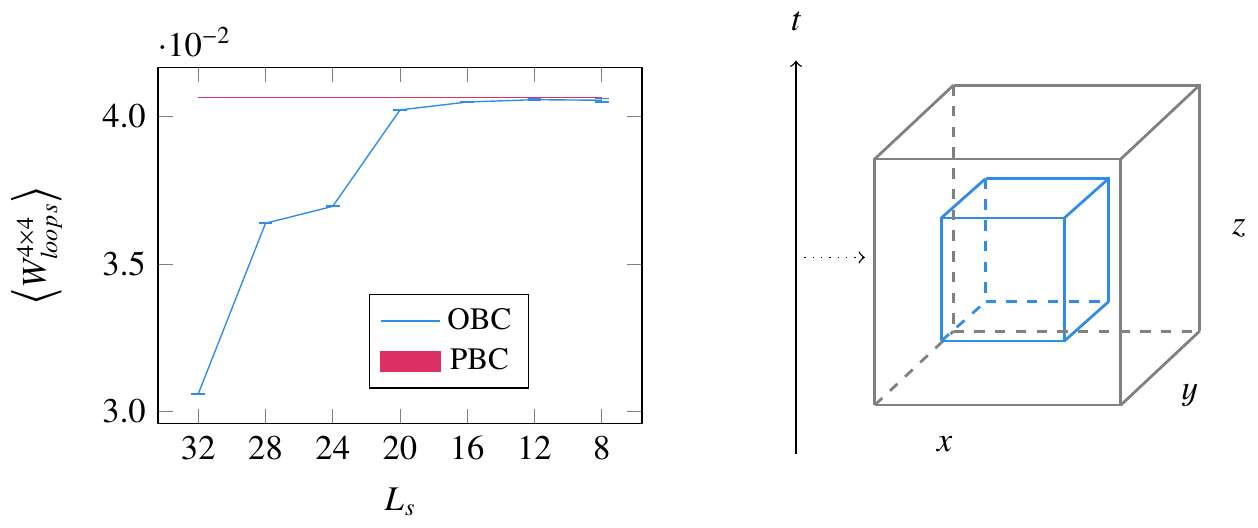}
				\caption{Finite size effects. We restrict the spatial extent to sub-cubes of volume $L_s^3$ (drawing on the right-hand side) and plot the expectation values of Wilson loops as a function of $L_s$. As expected, the OBC bulk results match the PBC expectation value. Note however the large extent where the finite size effects dominate.}
				\label{fig:finitesize}
			\end{center}
		\end{figure}

	This is of course not the whole story. The OBC come with their own disadvantages, the main one being the strength of finite size effects (for a careful discussion in the case of time-like OBC, the reader is referred to \cite{thiscontribSommer}). This problem is illustrated for the spatial OBC in figure~\ref{fig:finitesize}. There, as pictured on the right hand-side, we look at the average $4\times 4$ Wilson loops on sub-spatial volumes of $32^4$ lattices with OBC, $\beta=6.872$. As expected, the bulk value matches with the PBC value, where the finite-size effects are minimal. However, we observe that to get rid of them, we need to restrict the OBC configurations to $12^3\times 32$ sub-lattices, effectively losing a substantial part of the volume. This makes the use of small OBC lattices unattractive, while being still practical for larger lattices.
	A milder disadvantage of OBC is related to the topology. The determination of the topological susceptibility is made harder; one does not have anymore a clear stopping criterion for the Wilson flow (for PBC, it is given by the quantization of the topological charge). Moreover, it needs to be computed directly from the two point functions of the topological charge density, see  \cite{Bonati:2017woi} for example.
	
	In the following section, we present some exploratory results on how these problems might be overcome by using the OBC as an algorithmic tool.
	
	\section{Switching the Boundaries}\label{sec:switch}
	
\begin{figure}
		\begin{center}	
		\includegraphics{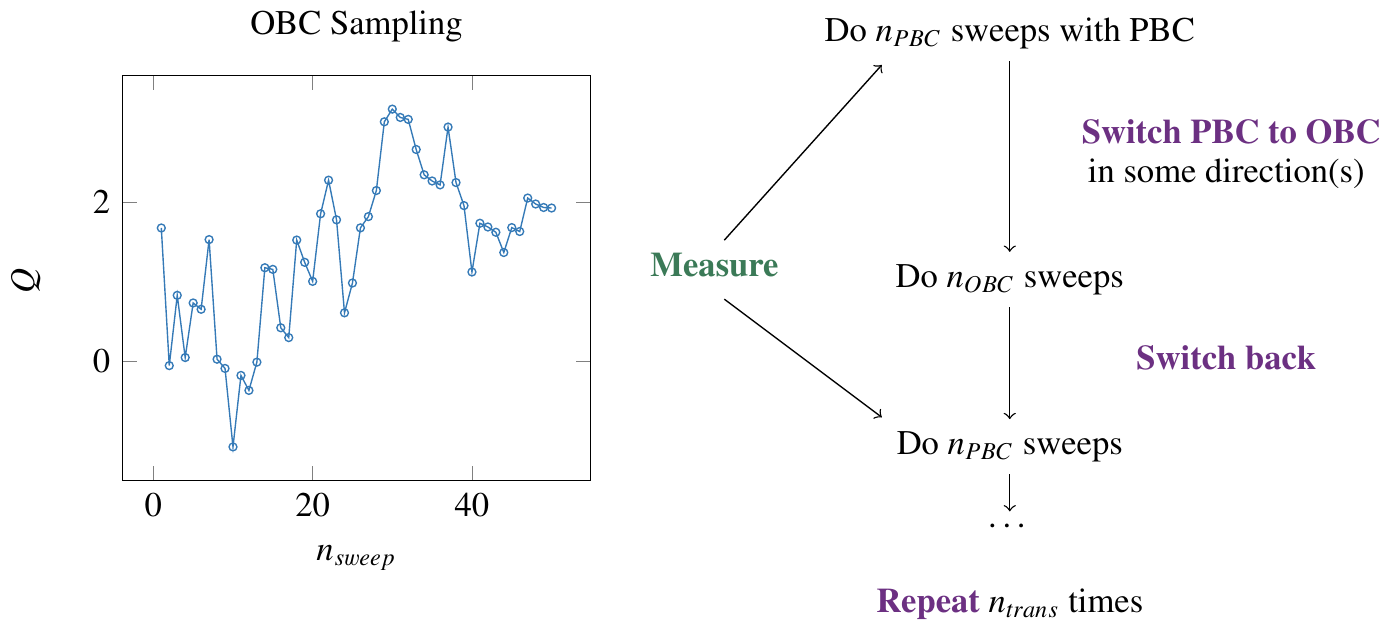}
			\caption{\textbf{Right: }Topological charge history during a MC run. Note that the $x$-axis represents the number of MC sweeps and \textit{not} a configuration number (configurations are separated by 500 sweeps). We see that the autocorrelation time of the topological charge with OBC is very short; the sampling is very efficient. \textbf{Left:} Illustration of how the "switch" algorithm works. Measurements are performed on periodic configurations. Re-thermalization is required after the OBC sweeps.}
			\label{fig:illustration}
	\end{center}
\end{figure}

	In order to cure the problems related to OBC, we try to combine them in the process of generating PBC configurations; we want to use them solely to "reshuffle" the topology. As we may see in figure~\ref{fig:illustration} plot, which depicts the  evolution of the topological charge with OBC as a function of MC \textit{sweeps}, OBC lead to a very efficient sampling of the topology. Our proposed algorithm is illustrated in the right-hand side of the same figure. We start a normal MC run with PBC. After $n_{PBC}/n_{sweeps}$ configurations, we switch PBC for OBC in one spatial direction, perform $n_{OBC}$ sweeps in order to change topological sector and then switch back to PBC. We repeat this $n_{trans}$ times, to have the required number of topological transitions.

\begin{figure}
			\begin{center}
		\includegraphics{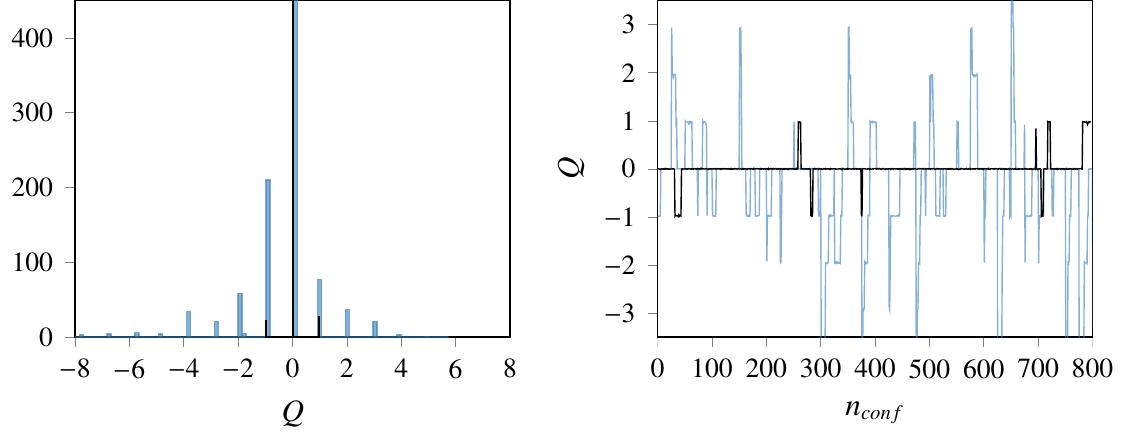}
		\caption{ Topological charge distribution, for the $64^3\times 16$ PBC (black) and "switch" (blue) configuration sets. The PBC results agree with the results of ~\cite{Borsanyi2016}; the "switch" algorithm seems to lead to some oversampling. Only re-thermalized configurations are shown. \textbf{Left:} Histogram. \textbf{Right:} MC-History.}
		\label{fig:switches}
			\end{center}
	\end{figure}

	We keep only the PBC configurations which can be considered to have re-thermalized (see the end of this section  for a discussion on this issue) after the OBC sweeps. Moreover, every time we proceed to a new set of OBC sweeps, we "open-up" a different direction, in order to homogenize the potential finite-size biases hence introduced. Our first results on $64^3\times 16$ lattices at $\beta=6.872$ ($1.28T_c$) are shown in figure~\ref{fig:switches}. The results obtained by doing the switches are displayed in blue. We see that an efficient sampling of the different topological sectors is achieved. It allows us to compute the topological susceptibility, to find $\frac{\chi_{switch}}{T_c^4}\approx 0.16 $. Actually, the sampling seems to be too efficient. The PBC results are shown in black.  The measured topological susceptibility $\frac{\chi_{PBC}}{T_c^4}\approx 0.0045$ is in agreement with the value of~\cite{Borsanyi2016}. The low sampling of the topology seems to  be a physical effect due to the damping of the quark-gluon plasma and does not seem to be a sign of topological freezing.
	
	This potential problem of oversampling is most certainly explained by the fact that brutally "opening-up" our lattice torus to allow the topological charge to flow in and out disturbs durably the systems. To reduce this issue, as mentioned before, we re-thermalized our configurations after every OBC sequence. 
	
	This operation is delicate for the following  reason: it is not an easy task to disentangle the finite size-effects related to fixed topology from spurious effects coming from badly thermalized lattices. To proceed, we looked at Wilson loops, computed on subset of configurations generated continuously after an OBC transition. We define re-thermalized lattices to be the ones which, in a part between two OBC sequences, lead to a set of Wilson loops expectations values which consistently comes from a single probability distribution. It is not completely clear how appropriate this procedure is; gaining a better understanding on the re-thermalization is a potential way of improving these results.
	
 As suggested to us during this conference (and following the idea of \cite{thisContriHasen}), a seemingly more rigorous way of building such an algorithm could involve  some parallel tempering between OBC and PBC configurations. This would remove the re-thermalization step and provide an algorithm whose properties are better defined. Implementing such an algorithm would also allow us to adjust or rule out our naive switch algorithm.

	\section{Conclusion}\label{sec:conclu}
	
	In lattice simulations, a good sampling of the different topological sectors is desirable, both for thermal and non-thermal simulations. In order to try to achieve such a sampling, we presented some results obtained with OBC imposed in spatial directions, having in mind to use them in the former case. We showed that, as expected, the topological charge is not quantized and has a very small auto-correlation time. We also showed that these conditions suffer from large finite-size effects. Building on these observations, we presented some exploratory results where we made use of OBC as a tool to change topological sectors between different PBC configurations. Our naive algorithm seems to lead to some oversampling. It is worth noting that even if it also leads to a wrong topological sampling, it has the merit of providing a new way to generate higher charge configurations, which could for example  be used in some re-weighting procedures.
	
	All in all, these first results are encouraging, but work remains to be done. First, the naive algorithm should be further adjusted, by gaining a better understanding of the re-thermalization procedure. Then, one should try to design a more subtle algorithm, using some parallel tempering methods for example. The next step would be to study fermionic configurations, one of the final objectives being to be able to use it as a different way to extract the topological susceptibility from lattice simulations.

	\subsection*{Acknowledgments}
	
	A.F. would like to thank A. Rothkopf, who kindly accepted to share his code~\cite{Rothkopf:2011db}. It was modified and used to generate configurations in the initial part of this project. O. K. and L. M. acknowledge support by the Deutsche Forschungsgemeinschaft (DFG) through the grant
	CRC-TR 211 "Strong-interaction matter under extreme conditions". The work of Y.B. and A.F.  was supported by the Swiss National Science Foundation. 

	\clearpage
	\bibliography{lattice2017}
	
\end{document}